# Single-photon interference experiment over 100 km for quantum cryptography system using a balanced gated-mode photon detector

Hideo Kosaka, Akihisa Tomita, Yoshihiro Nambu, Tadamasa Kimura, and Kazuo Nakamura

We demonstrate single-photon interference over 100 km using a balanced gated-mode photon detector and a plug & play system for quantum key distribution. The visibility with 0.1 photon/pulse was more than 80% after 100 km transmission. This corresponds to the fidelity of a quantum cryptography system of more than 90% and a QBER of less than 10%, satisfying the security criteria.

*Introduction:* Quantum cryptography or, more precisely, quantum key distribution (QKD) allows two remote parties (Alice and Bob) to generate a secret key, with privacy guaranteed by quantum mechanics [1-2]. Extensive efforts by numerous groups [3-12] has been devoted to extending the QKD distance using optical fibres ever since the introduction of the BB84 protocol by Bennet and Brassard in 1984 [1] and their first bench-top implementation of QKD over 30 cm of free-space in 1992 [2]. The first breakthrough was made by Townsend et al. in 1993 [3] using a phase modulator in a Mach-Zehnder interferometer instead of using polarization based systems. They achieved 10-km transmission of a single photon with high visibility, which was one-order longer transmission than that for polarization-based methods. The next breakthrough was made by using Faraday mirrors to self-align the polarization and to self-balance the path length of the interferometer. This was demonstrated by Muller et al. [9] and they called it the plug and play (P&P) interferometric system. The transmission length was at first limited to 23 km, but recently Stucki et al. have succeeded in extending QKD over 67 km using the P&P system [12].

To significantly extend the QKD distance, it is necessary to have better components. One of the key components for QKD is the single-photon detector (SPD), which is commonly made of avalanche photodiodes (APDs). The figure of merit for SPD is the extinction ratio, or S/N ratio, between the signal counts (S) and dark counts (N). By balancing two APDs, we have improved the S/N ratio by more than one order of magnitude compared to the conventional usage of APDs [13]. In this report, we outline the results of testing interference fringe visibility by combining the balanced, gated-mode photon detector and the P&P system over 100 km in conventional optical fibers.

*Balanced, Gated-mode photon detector:* InGaAs/InP-based APD is commonly used for photon detection at 1.55 μm in the Geiger mode, where the reverse bias higher than the breakdown voltage is applied. Since the Geiger mode operation results in rather high dark counts, pulsed bias synchronized to the detection time window is commonly used. This is called the gated mode. The gated mode operation in turn results in strong spikes on the transient signals and requires an unnecessarily high threshold in qubit discrimination. This spiking problem for gated-mode operation is resolved by taking the balanced output of APDs [13], which are usually used as a set for qubit discrimination. As a result, we can reduce the voltage for the pulsed gate as well as the threshold of the discriminator, resulting in lower dark counts while keeping the same detection quantum efficiency or single counts. The circuit diagram of the balanced, gated-mode photon detector is shown in Fig. 1. Using this method we achieved a dark count probability of $2 \times 10^{-7}$ at the detection quantum efficiency of 10% at -106.5°C.

*Experiment:* The experimental system is based on a phase-encoded interferometric QKD scheme [3] with a P&P configuration [9-12], as shown in Fig. 1. The system consists of 100-km-long fiber, a Mach-Zehnder interferometer on the Bob side and a Faraday mirror on the Alice side, with two phase-modulators on each side. The system operates at a wavelength of 1.55 μm and the laser light pulse is attenuated to get pseudo single-photon level transmission. The average photon number per pulse coming back from Alice was set to 0.1 in this experiment. The repetition rate for the laser operation, the APD gating, and biasing of the phase modulator was 500 kHz to avoid an after-pulse effect. Pulse widths for the laser, APDs and phase modulator were 0.5 ns, 0.75 ns, and 20 ns, respectively.

*Transmission characteristics:* We measured photon counting probability or key generation rate as a function of distance as shown in Fig. 2. The photon counting probability decreases almost exponentially with an increase in distance. Measured points are well fitted by a linear line with a transmission loss of the used fibre (0.25 dB/km). The interference fringe is also shown in the inset of Fig. 2. The interference fringe visibility $V = (I_{max}-I_{min})/(I_{max}+I_{min})$ after 100-km transmission was 83% and 80% for APD1 and APD2, respectively. These visibilities correspond to a fidelity of the quantum QKD system, defined as $F = (V+1)/2$, of more than 90% and a quantum bit error rate (QBER), defined as $1-F$, of less than 10%. A drift in temperature of the fibres and the resulting timing deviation during the measurements probably causes the deviation of the measured points from the fitted line that starts around 80 km. This can be eliminated by actively adjusting the gating time of the APDs. Temporal broadening of the photon arrival time due to a wavelength dispersion of the fibres, 1.7 ns/nm-100 km at 1.55 μm, must be negligible. Base lines in Fig. 2 show dark counts or detector noise (~ 0.1 count/s or $2 \times 10^{-7}$ per pulse) and error counts caused by stray photons (~ 0.6 count/s or $1.2 \times 10^{-6}$ per pulse). The detector noise corresponds to an S/N ratio of 57 dB. The stray photons are mainly caused by Rayleigh backscattering during fibre transmission, which appears after connecting the transmission line and still exists even when the Alice part is removed. Furthermore, the number becomes roughly a quarter by halving the repetition rate. We do not know the exact function of backscattering versus distance but it was roughly the same from 40 to 100 km. By using the balanced, gated-mode photodetector, we could increase the S/N ratio by 17 dB compared to a conventional unbalanced photodetector. Nevertheless, we could only extend the distance by an amount corresponding to 9 dB because of the backscattering noise. This backscattering will be reduced by inserting a storage line on the Alice side and using burst photon trains [11]. If we insist on using 10% as the QBER criteria to have a secure QKD system, the maximum distance without the backscattering effect is estimated to be 140 km in Fig. 2. The fibre we used for this experiment had a loss of 0.25 dB/km, which is rather high compared to the common spec for current fibre. If we could use fibres with 0.17-dB/km transmission loss, which is already commercially available, this SPD would make it possible to achieve 200-km-long QKD.

***Conclusion:*** We have demonstrated 100-km-long single-photon transmission in a P&P

system with a balanced, gated-mode photodetector. The obtained interference fringe after 100-km transmission showed a visibility of more than 80%, which corresponds to a fidelity of the quantum cryptography system of more than 90% and a QBER of less than 10%. This demonstration suggests that 100-km-long QKD is possible using our improved single-photon detector.

*Acknowledgment:*

This work was partly supported by the Telecommunications Advancement Organization of Japan.

**Authors' affiliations:**

Hideo Kosaka, Yoshihiro Nambu, and Kazuo Nakamura (Fundamental Research Laboratories, NEC, 34 Miyukigaoka, Tsukuba 305-8501, Japan)

Akihisa Tomita (Quantum Computation and Information Project, ERATO, JST, 34 Miyukigaoka, Tsukuba 305-8501, Japan)

Tadamasa Kimura (Department of Materials Science and Engineering, Tokyo Institute of Technology, 4259 Nagaoka, Yokohama 226-8502, Japan)

**Figure Captions**

Fig. 1. Schematic diagram of single-photon quantum cryptography system. FM: Faraday mirror, PM: phase modulator, PBS: polarization beam splitter, PC: polarization controller, A: attenuator, DS: discriminator, CT: counter

Fig. 2. Raw and net key generation rate as a function of transmission distance. An inset shows photon count rate as a function of phase modulation over 100 km

Figure 1

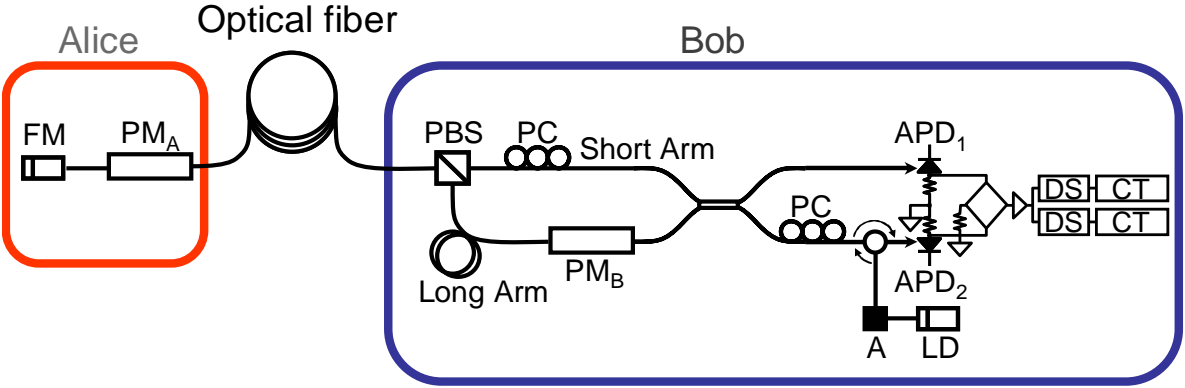

Figure 2

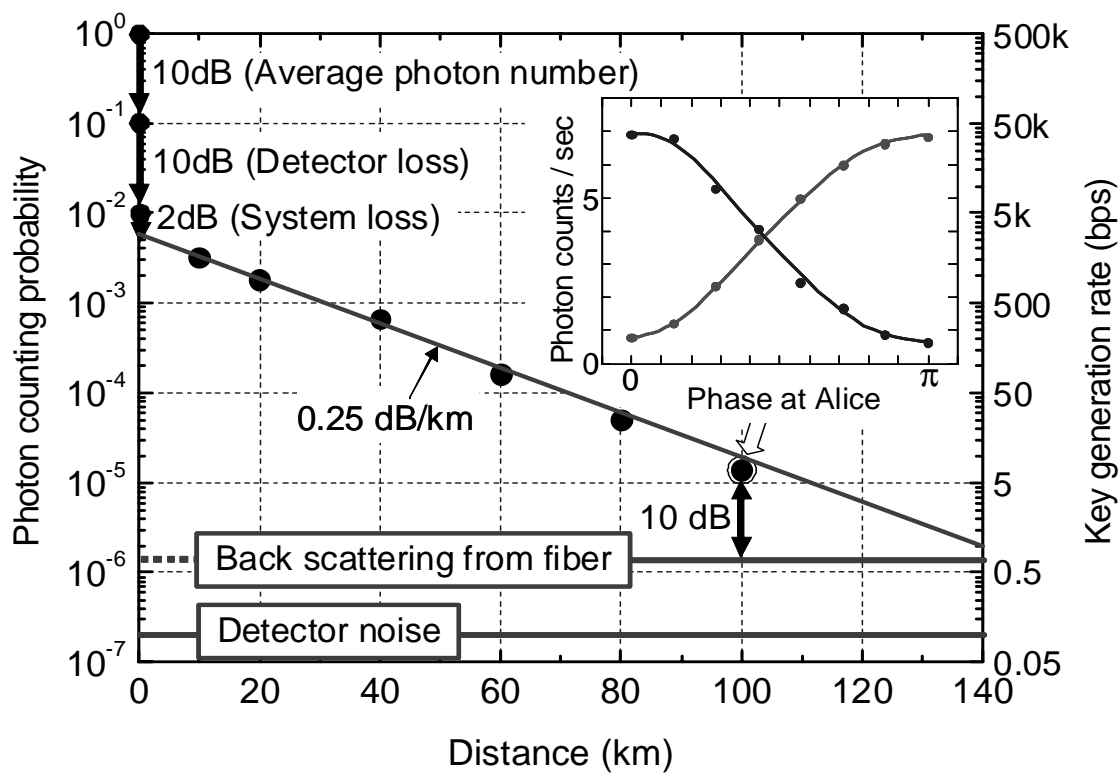